\documentstyle[12pt]{article}

\newcommand \cint {\int_{-\mbox{\scriptsize i}\infty}
                       ^{\mbox{\scriptsize i}\infty}}
\newcommand \rmi  {\mbox{i}}
\newcommand \tr  {\mbox{Tr}}
\newcommand \cN  {{\cal N}}
\newcommand \thb  {\bar{\theta}}
\newcommand \bthb {\bar{\mbox{\boldmath $\theta$}}}
\newcommand \bth  {\mbox{\boldmath ${\theta}$}}

\begin{document}
\title{Mean-Field Equations for Spin Models with Orthogonal
Interaction Matrices}
\author{Giorgio Parisi
        and Marc Potters\thanks{\tt potters@uranus.roma1.infn.it}
\\[0.5em]
  {\small Dipartimento di Fisica and INFN, Universit\`a di Roma
    {\em La Sapienza}}\\
  {\small \ \  P. A. Moro 2, 00185 Roma (Italy)}\\[0.5em]}
\date{March 1, 1995}
\maketitle

\begin{abstract}
We study the metastable states in Ising spin models with orthogonal
interaction matrices. We focus on three realizations of this model,
the random case and two non-random cases, i.e.\ the fully-frustrated
model on an infinite dimensional hypercube and the so-called
sine-model. We use the mean-field (or {\sc tap}) equations which we
derive by resuming the high-temperature expansion of the Gibbs free
energy. In some special non-random cases, we can find the absolute
minimum of the free energy.  For the random case we compute the
average number of solutions to the {\sc tap} equations. We find that
the configurational entropy (or complexity) is extensive in the range
$T_{\mbox{\tiny RSB}}<T<T_{\mbox{\tiny M}}$. Finally we present an
apparently unrelated replica calculation which reproduces the
analytical expression for the total number of {\sc tap} solutions.
\end{abstract}

\vfill

\begin{flushright}
  {\bf  cond-mat/9503009}
\end{flushright}

\newpage
The aim of this paper is to study the mean field equations (the {\sc
tap} equations) for the local magnetization for the fully frustrated
Ising model on an hypercube, or equivalently on a single cell of an
hypercubic lattice in the limit of infinite dimensions.

The model is very interesting and it has been widely studied.  It
belongs to the wide class of models which have a non-random
Hamiltonian but they behave in a similar (or identical) way to other
random models.  These models can be studied using the usual techniques
for random systems (e.g.\ the replica method) and in this way one can
obtain the correct results (with maybe the exception of the
equilibrium behavior at low temperature) \cite{BOUMEZ,MAPARA}.

In this model (as in the other models of the same class) there are at
least two transitions:
\begin{itemize}
\item The dynamical transition
at which the correlation time diverges.  At this temperature static
(i.e. equal time quantities) are fully regular.

\item An equilibrium transition at which the replica symmetry is
broken. Below this temperature there are many equilibrium states
available to the system
\end{itemize}

If we cool an infinite system below the dynamical transition, its
energy does not go to the equilibrium energy and the system remains
trapped in a metastable state of higher energy.

Generally speaking one can associate to each stable or metastable
state a solution of the {\sc tap} equation. On the contrary the
inverse relation does not hold. There are many solutions of the {\sc
tap} equations which do not correspond to stable of metastable
states. According to the folklore a {\sc tap} solution correspond to a
metastable state only if it is separated by other solutions by a
barrier whose height diverges with the volume; it seems most of the
solutions are separated by other solutions by finite barriers.

The relation between metastable states and the exponentially large
persistency time in a given metastable state from one side, and the
properties of the solutions of the {\sc tap} equations has never fully
understood. The aim of this paper contribute to this direction by
computing some of the properties of the solutions of the {\sc tap}
equation in a model which has metastable states.

An other interesting property of the model (absent in its random
correspondent) is the presence of a very low energy state, which
cannot be reached with continuity coming from the high temperature
phase.  At a given temperature the system has a first transition to a
state with very low energy (the crystal state).  The behavior of the
system in this low temperature phase may be understood with good
precision by considering the corresponding {\sc tap} equation.

There is also an other point on which we would like to call the
attention of the reader. For the Sherrington-Kirkpatrick ({\sc sk})
model it has already noticed that the computation of the total number
of solutions of the {\sc tap} equations turns out to be equivalent to
a apparently unrelated replica computation.  More precisely it was
found that:
\begin{equation}
\overline{\cN} \sim \lim_{m\to\infty} Z_m,
\end{equation}
where prefactors have been neglected and only the exponentially large
terms have been taken into account $Z_m $ is the partition function
computed by breaking the replica symmetry into two groups of replicas,
one with $m$ elements, the other with $n-m$ terms.

The relation is quite surprising, because the r.h.s.\ may be evaluated
without having to write down the very {\sc tap} equation.

Apparently there is no known explanation for this phenomenon. Our
contribution to this point is to observe that the previous relation
holds also in this case, where the form of the {\sc tap} equation is
much more complex. It is quite like (as also suggested by Cugliandolo
and Kurchan), that there should be an isomorphism of two algebraic
structure which explain this equality, but we have not done progresses
in this direction.

It may be interesting to recall that a similar phenomenon happens in
ferromagnetic systems with a random temperature or magnetic field.
Let us consider the case of random magnetic field.  Here we are
interested find the probability of having more than one solution to
the stochastic differential equation
\begin{equation}
-\Delta \phi(x) +m^2 \phi(x)+ g\phi(x)^3=h(x),
\end{equation}
which plays the role of the {\sc tap} equation in this model.

In the replica approach \cite{AMORPH,DISSIM,DOTPAR} it was found that
the existence of many solutions is related to the presence of non
trivial saddle points, where the field $\phi_a(x)$ for $a=1,n$ ($n=0$)
is of the form
\begin{eqnarray} \nonumber
\phi_a(x) =f(x) \ \ \mbox{for} \ \  1\le a\le m \\
 \phi_a(x) =f(x)+{g(x) \over m}
\ \ \mbox{for}  \ \  m< a\le n,
\end{eqnarray}
and one consider the limit where $m \to \infty$.

Also in this case the replica symmetry is broken by dividing the
replicas into two groups, one with $m$ elements, the other with $n-m$
terms, and taking the limit $m \to \infty$.  This phenomenon seems to
be quite widespread.

The plan of the paper is the following.

In section two we recall the definition of three different models,
which have the same high temperature expansion: the random orthogonal
model ({\sc rom}) where the coupling matrix is a random orthogonal
matrix, the sine model, where the elements of the coupling matrix can
be written as the sine of an appropriate expression, and the fully
frustrated models on a single hypercubic cell ({\sc ffm}). These three
models are identical in the high temperature expansion, because they
have very similar coupling matrices.

In section three we use the high temperature expansion to derive the
{\sc tap} equations.

In section four we study the property of the solution of the {\sc tap}
equation corresponding the lowest lying state (of zero energy) which
exists in the sine model and likely exists in the fully frustrated
hypercube (the so called {\it crystal} state).

In section five we compute the average number of {\sc tap} solution
for {\sc rom}. We study their properties as function of their free
energy both at $T=0$ and $T\ne 0$.  We also find a relation among the
properties of the {\sc tap} equations and the marginality condition
for the dynamical transition and show that the replica symmetric free
energy can be written as the sum of contributions from a large number
of metastable states above the {\sc rsb} transition.

In section six we present the computation of the total number of {\sc
tap} solutions with the two replica group method we have described
before.

Finally in the appendix we present some technical details  needed in
the computations of section five.

\section{The Model}

In what follows we will consider the model defined by the Hamiltonian
\begin{equation}
H=-{\textstyle \frac{1}{2}}\sum_{ij}J_{ij}\sigma_i\sigma_j,
\end{equation}
where $\{\sigma_i\}$ is a set of $N$ Ising spin variables
($\sigma_i=\pm 1$) and $J_{ij}$ is an $N\times N$ symmetric orthogonal
matrix with large connectivity ($z\gg 1$).  To lighten the notation,
the matrix $J_{ij}$ will be taken to have zeros on the
diagonal\footnote{Strictly speaking a generic matrix $\bf J$ in {\sc
rom} and the one for the sine model have non-zero diagonal elements,
nevertheless the matrix obtained by setting those elements to zero
will also be orthogonal in the large $N$ limit.}.

\subsection{Random Orthogonal Model}
In the random orthogonal model ({\sc rom}) the coupling matrix is
chosen at random in the set of orthogonal symmetric matrices. The
probability distribution (or integration measure) is defined by
writing $\bf J=ODO^{-1}$ with $\bf D$ a diagonal matrix composed of
$\pm 1$ and $\bf O$ a generic orthogonal matrix (not necessarily
symmetric) whose probability distribution is defined by the Haar
measure on the orthogonal group. We will make use of the identity
\cite{MAPARB}
\begin{equation}
\label{E_GINT}
\int {\cal D}{\bf J} \exp\left\{\tr\frac{{\bf J A}}{2}\right\}=
\exp\left\{N\tr\, G\left(\frac{{\bf A}}{N}\right)\right\},
\end{equation}
which holds in the large $N$ limit when ${\bf A}$ is a symmetric
matrix of finite rank and where $G(x)$ is given by
\begin{equation}
  \protect\label{E_GOFX}
  G(x)=- \frac{1}{4}\log(\frac{\sqrt{1+4x^2}+1}{2})+
  \frac{1}{4}\sqrt{1+4x^2}-\frac{1}{4}.
\end{equation}
Notice that integration over matrices chosen from a Gaussian
distribution ({\sc sk} model) also yields Eq.\ (\ref{E_GINT}) but with
$G(x)=x^2/4$.

\subsection{Sine Model}
The coupling matrix for the sine model is given by
\begin{equation}
\label{E_SINE}
J_{ij} = \frac{2}{\sqrt{2N+1}} \sin\left(\frac{2\pi i j}{2N+1}\right).
\end{equation}
The matrix $\bf J$ is obviously symmetric and its orthogonality follows
from orthonormality relations among simple harmonics. This model was
introduced in \cite{MAPARB} as a simple Hamiltonian that admits a
complex ground state for special values of $N$. More precisely, if
$2N+1$ is prime and $N$ is odd then the Legendre sequence
\begin{equation}
\sigma_j=j^N \pmod{2N+1},
\end{equation}
which consists of $\pm 1$ is a ground state configuration of the sine
model with energy density $-1/2$.  This fact is by no means obvious;
the interested reader is referred to \cite{MAPARB} for details.

Monte Carlo simulations have shown that the thermodynamical properties
of the sine model are the same as that of {\sc rom}.  The only
difference is that for the sine model with $2N+1$ prime there exists a
low-lying state---never seen for large $N$ when cooling from high
temperature---in which the system remains (at low temperature) if it is
put there by hand.  One explanation of the similarity between the sine
model and {\sc rom} is that Eq.\ (\ref{E_SINE}) can be viewed as a
(bad) pseudo-random generator, and therefore the couplings are, for all
practical purpose, random.

\subsection{Fully-Frustrated Model}
Frustration in an Ising-like system (with $J_{ij}=\pm 1$) is defined
by the product of the couplings over a given plaquette \cite{TOUL}.
If this product is $-1$ the plaquette is said to be frustrated. One
can construct a coupling matrix on a $d$ dimensional simple cubic
lattice such that every plaquette is frustrated. This construction is
not unique but all realizations are gauge equivalent so the
thermodynamical properties of the model are unique. For a single
hypercubic cell, the fully-frustrated condition imposes \cite{DPTV1}
\begin{equation}
\sum_k J_{ik}J_{kj}= d\, \delta_{ij}.
\end{equation}
Therefore if we divide the couplings
by $\sqrt{d}$ the matrix $\bf J$ will be orthogonal and symmetric.
The thermodynamical limit will be taken by letting $d$ go to infinity,
which will also insure that the coordination number is large.

Note that unlike {\sc sk} on a cubic lattice, for the fully-frustrated
model ({\sc ffm}) the $d\rightarrow\infty$ limit of a single hypercube
is not completely equivalent to that of the lattice.  The reason is
that for {\sc sk} the distribution of the eigenvalues of the coupling
matrix (the celebrated Wigner semi-circle law) is the same both on the
single cell and on the lattice.  In the fully-frustrated case, the
matrix $\bf J$ has only two eigenvalues $\{-1,1\}$ for the unit
hypercube, while it has a continuous distribution for the lattice, in
the range $[-2,2]$ with rms value $\sqrt{2}$ \cite{DPTV2}.
Heuristically, in the high-T phase the system is sensitive to all
eigenvalues of the matrix: as we shall see, both models (lattice and
unit cell) have the same free energy up to a rescaling of $\beta$ by a
factor of $\sqrt{2}$.  On the other hand, in the low-T phase, it is
the largest eigenvalues of $\bf J$ that matter: for example the ground
state of the lattice has twice the energy density of that of the unit
cell.  It is this difference between $2$ and $\sqrt{2}$ that makes it
impossible to go from a single cell to the lattice with a simple
rescaling of the temperature: their qualitative behavior should
nevertheless be the same.

The ground state configuration of {\sc ffm} is not known for $d>8$.
It is conjectured that, with the $\sqrt{d}$ normalization, the ground
state energy density goes to $-1/2$ for large $d$ and that it is equal
to $-1/2$ when $d$ is a perfect square.  As with the sine model, Monte
Carlo simulations of {\sc ffm} are well described by {\sc rom};
furthermore they show aging effects, which strengthens our belief that
{\sc ffm} is a glassy system \cite{MAPARFF}.

\section{Mean-Field Free Energy}
\subsection{Standard High-Temperature Expansion}
Our goal is to derive the {\sc tap} equations for the orthogonal model
using the high-temperature expansion of the magnetization-dependent
(Gibbs) free energy.  Before doing so, let us try to understand the
standard ($m_i=0$) high-temperature expansion.  We can write
\begin{equation}
\label{E_FREEPHI}
e^{-\beta F}=
2^{N}
\mbox{det}^{1/2}(\beta{\bf J})
\int\prod_i \frac{d\phi_i}{\sqrt{2\pi}}
\exp\left\{-\frac{1}{2 \beta}\sum_{ij}J^{-1}_{ij}
\phi_i\phi_j +\sum_i \log\cosh(\phi_i + h_i)\right\},
\end{equation}
where we have introduced a site-dependent magnetic field $h_i$ for
later convenience; in this section we set it to zero.  We can view
this partition function as a theory of $N$ fields in zero dimension.
In this language the high-temperature free energy $F(\beta)$ is the
sum of connected diagrams whose vertex factors are the coefficients of
the Taylor series for $\log\cosh x$, with the propagator $\beta
J_{ij}$.  The first few terms are
\begin{eqnarray} \nonumber
\label{E_DIAG}
-\beta F(\beta) &=& N\log 2
+\frac{\beta^2}{4} \sum_{ij} J^2_{ij}
+\frac{\beta^3}{6} \sum_{ijk} J_{ij}J_{jk}J_{ki} \\
&&\mbox{}-\frac{\beta^4}{4} \sum_{ijk} J^2_{ij}J^2_{jk}
+\frac{\beta^4}{12} \sum_{ij} J^4_{ij}
+\frac{\beta^4}{8} \sum_{ijkl} J_{ij}J_{jk}J_{kl}J_{li}+O(\beta^5),
\end{eqnarray}
where $J_{ii}=0$ excludes `petals'---terms involving $\sum J_{ii}$.
The only terms that survive the large $N$ limit are the so-called
`cactus' diagrams---trees made out of loops joined at the vertices
(see Fig.\ 1).  The sum of $J_{ij}$'s for cactus diagrams is equal to
$N$ if the lengths of all loops are even (even-cacti) and zero
otherwise.  To show this, we take a cactus diagram and contract each
loop to a point starting from the outermost loops, the contraction is
done using the orthogonality relation
\begin{equation}
\sum_k J_{ik}J_{kj}=\delta_{ij}.
\end{equation}

In other words, the high-temperature free energy is given by $N$ times
the sum over the combinatorial factors and powers of $\beta$ of all
even-cactus diagrams.  In particular this free energy is independent
of the particular choice of the orthogonal matrix $\bf J$.

There is more than one way to calculate this sum.  Since it is
independent of the matrix $\bf J$ it must be equal to the (annealed)
average over such matrices, recovering the result of Ref.\ \cite{MAPARB}
\begin{equation}
\label{E_HFREE}
-\beta F(\beta) =N\log 2 +N G(\beta),
\end{equation}
with $G(x)$ given by Eq.\ (\ref{E_GOFX}).

More directly, we notice that the desired diagrams are precisely those
summed by the high-temperature series of the spherical model. In the
spherical model the Ising variables $\sigma_i$ are replaced by
continuous variables $S_i$ constrained to have $\sum_i S_i^2 =N$.
Fixing this last constraint with a Lagrange multiplier, the sum over
all configurations is then a Gaussian integral which can be done
exactly knowing the eigenvalues of the matrix $\bf J$, i.e.,$1$ and
$-1$ in equal proportions. After eliminating the Lagrange multiplier,
we find
\begin{equation}
\label{E_HFREESP}
-\beta F_{\mbox{\scriptsize sph}}(\beta) =\frac{N}{2}(\log 2\pi+1)
+N G(\beta),
\end{equation}
with the same $G(x)$. The first terms in Eq.\ (\ref{E_HFREE}) and Eq.\
(\ref{E_HFREESP}) are the volume of phase space of the two respective
models.  A similar analysis was done in Ref.\ \cite{DPTV2} for {\sc
ffm} on the full lattice. They also found Eq.\ (\ref{E_HFREE}) but
with $G(\beta)$ replaced with $G(\sqrt{2}\beta)$.

For comparison, the high-temperature series for {\sc sk} is easily
done; the only surviving diagram is the first one in Eq.\
(\ref{E_DIAG}) leaving us with
\begin{equation}
-\beta F_{\mbox{\tiny SK}}(\beta)= N\log 2 + N\frac{\beta^2}{4},
\end{equation}
where we have used that $\sum J_{ij}^2$ is self-averaging.

\subsection{Gibbs Free Energy}
The high-T free energy we have just calculated does not tell us
anything about the low-temperature phase.  It does not have any
singularity for positive $\beta$ which might signal a phase
transition.  What we need to compute is the magnetization-dependent
free energy (often called the Gibbs free energy).  In {\sc sk} this
quantity is exactly the {\sc tap} free energy which is still valid in
the low-temperature phase.  The program is simple, we need to expand
Eq.\ (\ref{E_FREEPHI}) in powers of $\beta$ and $h_i$ and perform a
Legendre transform, passing from the variables $h_i$ to $m_i$
using
\begin{equation}
\Phi(\beta,m_i)\equiv F(\beta,h_i)-\sum_i h_i m_i
\quad\mbox{ with }\quad
m_i\equiv\frac{\partial F}{\partial h_i}.
\end{equation}
Calculating the first few terms in $\beta$ by hand, we find
\begin{eqnarray} \nonumber
-\beta\Phi(\beta,m_i)&=&
-{\textstyle \frac{1}{2}}
\sum\left\{(1+m_i)\log\left[{\textstyle \frac{1}{2}}(1+m_i)\right]
+(1-m_i)\log\left[{\textstyle \frac{1}{2}}(1-m_i)\right]\right\}\\
&&+\frac{\beta}{2}\sum_{ij} J_{ij} m_i m_j
+\frac{\beta^2}{4}\sum_{ij}J^2_{ij}(1-m^2_i)(1-m^2_j)
+O(\beta^3).
\end{eqnarray}
The first term is the entropy of Ising spins constraint to have
magnetization $\{m_i\}$, the second one is minus the `na\"{\i}ve'
mean-field energy, and the third corresponds to the Onsager reaction
term. In Ref.\ \cite{PLEFKA} the {\sc tap} equations for {\sc sk} were
re-derived in this way, the higher order terms being negligible in
that case.  The analysis of the previous section ($m_i=0$) tells us
that for a model with an orthogonal matrix, the high-temperature
series will include an infinite number of non-negligible terms. We
must therefore find a systematic way of calculating all the terms in
the Gibbs free energy.

In standard field theory, the Legendre transform can be formulated in
diagrammatical terms.  The Gibbs free energy, also called the
effective potential in particle physics, is given by the sum over
connected one-particle-irreducible diagrams.  This is not true in our
case.  The reason is that the external field $h_i$ does not appear as
a linear source term in Eq.\ (\ref{E_FREEPHI}), instead it appears
inside the potential.  A shift in $\phi_i$ would not simplify matters
since it would introduce a term quadratic in $h_i$ sufficient to
render invalid the usual 1PI derivation.  The present authors don't
know of any diagrammatic expansion of the Gibbs free energy for
Ising-like system.

It is however possible to perform such an expansion algebraically and
to give it a diagrammatic interpretation \cite{YEDGE2}.  The weak
point of this method is that the vertex weight and the combinatorial
factors cannot be calculated systematically.  Nevertheless, one can
see that the cactus diagrams\footnote{What we call cactus diagrams are
called `loop diagrams' in Ref.\ \cite{YEDGE2}.  The difference in
terminology arises from the fact that these authors consider
restricted sums (sums over different indices with combinatorial
factors) while we consider unrestricted sums.  The vertex weights
coming from the expansion of $\log\cosh x$ are precisely those
necessary to go from one type of sum to the other.}, will appear in
the Gibbs free energy with the same weight but with an extra factor of
($1-m^2_{i_k}$) for each vertex $i_k$ \cite{YEDGE1}.  Therefore we
have
\begin{eqnarray} \nonumber
\label{E_BIGSER}
-\beta\Phi&=&
-{\textstyle \frac{1}{2}}
\sum\left\{(1+m_i)\log\left[{\textstyle \frac{1}{2}}(1+m_i)\right]
+(1-m_i)\log\left[{\textstyle \frac{1}{2}}(1-m_i)\right]\right\}
+\frac{\beta}{2}\sum_{ij} J_{ij} m_i m_j \\\nonumber
&&\mbox{}+\frac{\beta^2}{4}\sum_{ij}J^2_{ij}(1-m^2_i)(1-m^2_j)
-\frac{\beta^4}{4} \sum_{ijk} (1-m_i^2)J^2_{ij}(1-m_j^2)J^2_{jk}
(1-m_k^2)\\
&&\mbox{}+\frac{\beta^4}{8} \sum_{ijkl} (1-m_i^2)
J_{ij}(1-m_j^2) J_{jk}(1-m_k^2) J_{kl}(1-m_l^2)J_{li}
+O(\beta^6).
\end{eqnarray}

We now argue that in the large $N$ limit, all those terms (except for
the `entropic' and `energetic' one) are self-averaging, that is, the
error introduced by substituting $m_k^2$ by $q\equiv N^{-1}\sum_i
m^2_i$ vanishes in the thermodynamical limit.  Equivalently in {\sc
sk} the Onsager term in the free energy is often replaced by
$\beta^2(1-q)^2/4$.  We can now resum Eq.\ (\ref{E_BIGSER}), by
noticing that we are now summing over the same diagrams than in the
$m=0$ case.  We recover the reaction term of Eq.\ (\ref{E_HFREE}) with
$\beta$ replaced by $\beta(1-q)$,
\begin{eqnarray} \nonumber
\label{E_TAPFREE}
\beta \Phi&=&
{\textstyle \frac{1}{2}}
  \sum\left\{
     (1+m_i)\log\left[
                           {\textstyle \frac{1}{2}}(1+m_i)
                        \right]
     +(1-m_i)\log\left[
                           {\textstyle \frac{1}{2}}(1-m_i)
                         \right]\right\}\\
&&-\frac{\beta}{2} \sum_{ij} J_{ij} m_i m_j
-N G(\beta(1-q)),
\end{eqnarray}
where the function $G(x)$ is again given by Eq.\ (\ref{E_GOFX}).

The same result can be obtained in a more heuristic way.  We saw in
the previous section that the spherical model reproduces the
high-temperature expansion of the model with Ising spins except for
the entropic term.  It is quite plausible that the two models will
have the same Gibbs free energy, except once again for the entropic
term.  This argument has been used (the other way around) to deduced
the right {\sc tap} equations for the spherical p-spin model
\cite{KUPAVI}.  The general Gibbs free energy for a spherical model
with quadratic interaction is given by
\begin{equation}
\beta \Phi_{\mbox{\scriptsize sph}}=
\log \mbox{det}^{1/2}(\lambda-\beta{\bf J})-\frac{N}{2}(\log 2\pi
+\lambda)
+\frac{1}{2}\sum_{ij} m_i (\lambda \delta_{ij} -\beta J_{ij}) m_j,
\end{equation}
with $\lambda$ determined by the saddle point equation.  Specializing
to the case where the eigenvalues of $\bf J$ are $\{+1,-1\}$ in equal
proportions and after eliminating $\lambda$ we find
\begin{equation}
\beta \Phi_{\mbox{\scriptsize sph}}=-\frac{N}{2}(\log [2\pi(1-q)]+1)
-\frac{\beta}{2} \sum_{ij} J_{ij} m_i m_j -N G(\beta(1-q)),
\end{equation}
with $q\equiv N^{-1}\sum_i m^2_i$.  The first term is the entropy of
spherical spins constrained to have magnetization $m_i$. As expected,
it is the only term that differs from Eq.\ (\ref{E_TAPFREE}). Note
that in this case we did not need to substitute $1-m_k^2$ with its
average value ($1-q$).

Armed with the free energy (\ref{E_TAPFREE}), we can now write down
the `mean-field' or {\sc tap} equations for our model. They are given
by the partial derivatives of $\Phi$ with respect to the $m_i$'s:
\begin{equation}
\label{E_TAP}
\tanh^{-1} m_i +2 \beta G'(\beta(1-q))m_i-\beta\sum_{j}J_{ij}
m_j=0.
\end{equation}

We note finally that if we substitute in Eqs.\ (\ref{E_TAPFREE},
\ref{E_TAP}) $G(x)=x^2/4$ we recover the standard result of \cite{TAP}
for {\sc sk} just as the same substitution in our integration formula
(\ref{E_GINT}) recovers the integration over a Gaussian distribution.
In Sect.\ \ref{S_NUMB}, this fact will provide us with an easy way to
compare our formulae against those of Ref.\ \cite{BM}.

\section{Minimum of the Free Energy}

The orthogonality of $J$ imposes a simple bound on the energy of any
spin configuration. Indeed, a configuration vector $\{\sigma_i\}$ can
be decomposed into its projection $\{s_i^+\}$ on the eigenspace with
eigenvalue $+1$ and $\{s_i^-\}$ on the complement with eigenvalue
$-1$. Then we have
\begin{equation}
E({\bf\sigma}) = -\frac{1}{2}\sum_{ij} J_{ij} \sigma_i \sigma_j
=-\frac{1}{2}\left(\left|{\bf s}^+\right|^2
- \left|{\bf s}^-\right|^2\right).
\end{equation}
In other words, the energy is bounded below by $-N/2$ and this bound
is reached iff ${\bf\sigma}$ is an eigenvector of $\bf J$ with
eigenvalue $+1$.  We shall argue later that for a generic large
orthogonal matrix the existence of an eigenvector whose every
component is either $+1$ or $-1$ is highly improbable.  Nevertheless,
we can construct such a matrix, for example the sine model for odd $N$
with $2N+1$ a prime number.  Whether or not {\sc ffm} admits such a
ground state for special values of $d$ and/or in the limit
$d\rightarrow\infty$ is still an open question.  Unfortunately,
neither an $e=-1/2$ spin configuration for $d\geq 9$ has been found
nor has a proof that it cannot exist.

For the remaining part of this section, we will consider a model that
admits an $e=-1/2$ ground state. The {\sc tap} equations (\ref{E_TAP})
admit in this case a solution of the form
\begin{equation}
\label{E_ANSATZ}
m_i=\sqrt{q} \epsilon_i ,
\end{equation}
where the $\{\epsilon_i\}$ are $\pm1$ and form an eigenvector of $\bf
J$ with eigenvalue $+1$.  With this ansatz Eq.\ (\ref{E_TAP}) reduces
to
\begin{equation}
\label{E_LOWFQ}
q=\tanh^2\left\{
      \beta\sqrt{q}\left[
                 1+\frac{1-\sqrt{1+4\beta^2(1-q)^2}}{2\beta(1-q)}
      \right]
\right\},
\end{equation}
where we have used Eq.\ (\ref{E_GOFX}) to compute $G'(x)$. This
solution has specific free energy given by
\begin{equation}
\beta f=
    \frac{1+\sqrt{q}}{2}\log\left[
                           {\textstyle \frac{1}{2}}(1+\sqrt{q})
                        \right]
     +\frac{1-\sqrt{q}}{2}\log\left[
                           {\textstyle \frac{1}{2}}(1-\sqrt{q})
                         \right]
-\frac{\beta}{2} q
- G(\beta(1-q)).
\end{equation}
For a given value of $q$, this is the lowest free energy solution.  In
fact, the energetic term reaches its lower bound from orthogonality,
the entropy is maximum (at fixed $q$) when all the local
magnetizations are equal in magnitude and the reaction term only
depends on $q$.  The absolute minimum of the free energy must
therefore be of this form.  Eq.\ (\ref{E_LOWFQ}) always admits a
solution with $q=0$.  At low temperature it will also admit a solution
with non-zero $q$.  Numerically one finds that below
$T=0.400$\footnote{This paper contains many numbers obtained from
numerical analysis, they are noted with equal signs and every digit is
significant.}  a solution with $q=0.92$ appears but with higher free
energy that the paramagnetic solution.  At $T=0.178$ ($q=0.99995$)
this solution becomes the true minimum of the free energy.

One might conclude as in \cite{YEDGE1}, that the system undergoes a
first order transition at $T=0.178$. This transition is not seen in
Monte Carlo simulations of {\sc ffm} or the sine model when cooling
down from high temperature. Besides Eq.\ (\ref{E_ANSATZ}), there are
many other solutions to the mean-field equations. As we shall we in
the next section, for a generic interaction matrix and below a certain
critical temperature, their number grows exponentially with the size
of the system. It is those solutions and the large free energy
barriers between them that prevent the system from finding the true
minimum of the free energy and gives it instead a glassy behavior.

When the ground state configuration is known---as in the sine model
for special $N$---it is however possible to see this `crystalline
state' in Monte Carlo simulations.  One has to start in the ground
state at $T=0$ and continuously increase the temperature.  Fig.\ 10 of
Ref.\ \cite{MAPARB} shows the result of such a simulation (sine model
$N=44$ and $N=806$).  Calculation of the internal energy for our
solution reproduces exactly those curves.  Note that for the larger
value of $N$ the system stays trapped beyond the point where the
paramagnetic solution becomes stable ($T=0.712$ on their scale) this is
a clear sign that barriers between different metastable states are
very high.

The solution (\ref{E_ANSATZ}) can be transposed directly to the {\sc
ffm} on the lattice by taking care of the appropriate factors of $2$
and $\sqrt{2}$ recovering the result of Ref.\ \cite{YEDGE1}.  One
important point, though, is that this solution is only valid if the
ground state energy density is $-1/2$ ($-1$ for the full lattice),
otherwise $\{\epsilon_i\}$ is not an eigenvector of $\bf J$ and Eq.\
(\ref{E_TAP}) does not reduce to Eq.\ (\ref{E_LOWFQ}).

\section{Number of TAP Solutions}
\label{S_NUMB}
\subsection{General Result}
In this section we will compute the average number of metastable
states for {\sc rom}.  We will follow as much as possible the steps
and the notation of Ref.\ \cite{BM} where the corresponding result for
{\sc sk} was first derive. Recall our {\sc tap} equations:
\begin{eqnarray} \nonumber
\label{E_TAP2}
\Gamma_i&\equiv&\tanh^{-1} m_i +2 \beta G'(\beta(1-q))m_i
-\beta\sum_{i\neq j}J_{ij} m_j=0\\
&\equiv&g(m_i)-\beta\sum_{i\neq j}J_{ij}
m_j=0,
\end{eqnarray}
with corresponding free energy written as a sum of single site terms:
\begin{equation}
f=(\beta N)^{-1}\sum_{i}\left[-\log 2-G(\beta(1-q))-\beta
qG'(\beta(1-q))+
{\textstyle \frac{1}{2}} m_i \tanh^{-1} m_i
+{\textstyle \frac{1}{2}} \log (1-m_i^2) \right].
\end{equation}
We write the number of solutions as the integral over all possible
values of $m_i$ fixing with $\delta$-functions that the $\{m_i\}$ form
a solution of Eq.\ (\ref{E_TAP2}) with free energy $f$. Using a
Fourier representation of the $\delta$-function, we obtain
\begin{eqnarray} \nonumber
\label{E_NSOL1}
\cN_s (f)&=&N^2 \int_0^1 dq \cint \frac{d\lambda}{2\pi\rmi}\cint
\frac{du}{2\pi\rmi}
\cint\prod_i \left(\frac{dx_i}{2\pi\rmi}\right)
\int_{-1}^1\prod_i\left(dm_i\right)
\exp\left[-N(\lambda q+uf)\right.\\
&&\left.-\lambda\sum_i m_i^2 + u\sum_i f(m_i)
+\sum_i x_i g(m_i) -\beta \sum_{i<j}J_{ij} \left(x_i m_j+x_j m_i \right)
 \right] \left|\det {\bf A}\right|,
\end{eqnarray}
where
\begin{eqnarray} \nonumber
A_{ij}&=&\frac{\partial \Gamma_i}{\partial m_j}=\left[(1-m_i^2)^{-1}
+2 \beta G'(\beta(1-q))\right]\delta_{ij}-\beta J_{ij} \\
&\equiv&a_i \delta_{ij}-\beta J_{ij},
\label{E_ADEF}
\end{eqnarray}
dropping a term in $m_im_j/N$ (see appendix).

We can now proceed to average Eq.\ (\ref{E_NSOL1}) over the random
couplings $J_{ij}$. We should really be averaging $\log\cN_s$, the
extensive quantity, not $\cN_s$. To do so we would need to introduced
replicas.  The replica symmetric computation would be tedious but
straightforward leaving us not with 7 unknown parameters (as will be
the case below) but on the order of 14. On the other hand, the direct
average will be sufficient to provide us a clear picture of the
metastable states.

There are two terms that depend on the $\{J_{ij}\}$.  It is shown in
appendix that the two can be averaged independently.  Using Eq.\
(\ref{E_GINT}) to average the term involving $x_i$ and $m_i$, we find
\begin{eqnarray} \nonumber
\label{E_COUPINT}
\left\langle\exp\left[-\beta\sum_{i<j} J_{ij}\left(x_i m_j+x_j m_i
\right)\right]\right\rangle&=&
\exp\left\{N\tr G\left[
\beta\frac{{\bf x}\otimes{\bf m}+{\bf m}\otimes{\bf x}}{N}
\right]\right\}\\
&=&\exp\left\{N\left[G(v+\sqrt{w})+G(v-\sqrt{w})\right]\right\},
\end{eqnarray}
where
\begin{equation}
\label{E_VANDW}
v=\frac{\beta}{N}\sum_i x_i m_i\mbox{ and }
w=\frac{\beta^2 q}{N}\sum_i x_i^2 .
\end{equation}
The second equality in Eq.\ (\ref{E_COUPINT}) follows from an analysis
of the eigenvalues of the matrix ${\bf x}\otimes{\bf m}+{\bf
m}\otimes{\bf x}$.  This matrix has only two non-zero eigenvalues
corresponding to the two terms: $v+\sqrt{w}$ and $v-\sqrt{w}$.

To compute the average the determinant we will need to drop the
absolute value.  This corresponds to weighting each solution with the
sign of the determinant of its Hessian matrix.  Formally, we would be
computing a topological invariant (from Morse theory) which as little
to do with our original goal.  Nevertheless, the calculation without
the absolute value gives sensible results (here and in Refs.\
\cite{BM,RIEGER}) and connects smoothly with the zero
temperature results where the calculation can be done without this
pathology.  This problem is discussed in more detail in Refs.\
\cite{PARSOU,KURABS}.

We introduce a set of anti-commuting (Grassman) variables
$\{\theta_i,\bar{\theta_i}\}$ to express the determinant as an
exponential,
\begin{eqnarray} \nonumber
\langle\det {\bf A}\rangle
&=&
\left\langle\int d\bth d\bthb
\exp\left[\sum_i a_i \thb_i\theta_i
-\beta\sum_{(ij)}J_{ij}(\thb_i\theta_j+\thb_j\theta_i)\right]
\right\rangle\\
&=&\int d\bth d\bthb \exp\left\{\sum_i a_i \thb_i\theta_i+
N\tr G\left[
\beta\frac{\bthb\otimes\bth-\bth\otimes\bthb}{N}
\right]\right\} .
\end{eqnarray}
If we expand $\tr G[\ldots]$ in a Taylor series, we can
use the following propriety of Grassman variables
\begin{equation}
(\thb_{i_1}\theta_{i_2}+\thb_{i_2}\theta_{i_1})
(\thb_{i_2}\theta_{i_3}+\thb_{i_3}\theta_{i_2})
\ldots
(\thb_{i_n}\theta_{i_1}+\thb_{i_1}\theta_{i_n})
=-2\prod_{k=1}^n \thb_{i_k}\theta_{i_k},
\end{equation}
which allows us to resum the series and find
\begin{equation}
\label{E_FERM}
\langle\det {\bf A}\rangle=
\int d\bth d\bthb \exp\left\{\sum_i a_i \thb_i\theta_i
-2N G\left[\frac{\beta}{N}\sum_i \thb_i \theta_i\right]\right\} .
\end{equation}
Finally we fix $r=(\beta/N)\sum_i \thb_i\theta_i$ using a Lagrange
multiplier $R$, which allows us to perform the integration over the
Grassman variables, leaving us with
\begin{equation}
\label{E_DETFINAL}
\left\langle \det {\bf A} \right\rangle = \int_{-\infty}^{\infty}dr
\cint\frac{dR}{2\pi\rmi}\prod_i \left(a_i+\beta R\right)
\exp\{N[-rR-2G(r)]\}.
\end{equation}
We now collect results form Eqs.\ (\ref{E_NSOL1}), (\ref{E_COUPINT})
and (\ref{E_DETFINAL}), introduce Lagrange multipliers $V$ and $W$ to
impose Eqs.\ (\ref{E_VANDW}), perform the $x_i$ integration, drop
multiplicative prefactors and finally set $V=2G'(\beta(1-q)) -\Delta$,
$R=B/\beta-2G'(\beta(1-q))$ and $r=\beta b$, to obtain
\begin{eqnarray} \nonumber
\label{E_NFINAL}
\langle\cN_s (f)\rangle&=&\mbox{saddle }
\exp\left\{ N \left[-\lambda q -uf-bB-v\Delta -wW/q
+G(v+\sqrt{w})
\right.\right.\\
&&\left.\left.
\mbox{}+G(v-\sqrt{w})-2G(\beta b)+2(\beta b-v)G'(\beta(1-q))+\log I
\right] \right\}  ,
\end{eqnarray}
where
\begin{equation}
\label{E_INT}
I=\int_{-1}^1 \frac{dm}{2\beta\sqrt{\pi W}}
\left(\frac{1}{1-m^2}+B\right)
\exp\left[-\frac{\left(\tanh^{-1}m-\beta\Delta m\right)^2}{4\beta^2W}+
\lambda m^2
+uf(m)\right] .
\end{equation}
We have indicated by the term {\em saddle} that the right-hand side of
Eq.\ (\ref{E_NFINAL}) is to be extremized with respect to the nine
variables: $\lambda, q, u, b, B, v, \Delta, w, W$.  By partial
differentiation with respect to those variables, we obtain the saddle
point equations. They admit a solution with $B=0$ which implies
$b=1-q$; following Ref.\ \cite{BM} we adopt this solution.  This
choice will also lead us to the correct $T=0$ result.

As advertised earlier, the result for {\sc sk} can be obtained from
Eq.\ (\ref{E_NFINAL}) by setting $G(x)=x^2/4$.  In this case, three
variables can be eliminated using the saddle point equations to give
back the same expression as in Ref.\ \cite{BM}.

\subsection{$T=0$ Result}
Before looking at the numerical solution for the saddle point of Eq.\
(\ref{E_NFINAL}), let us study its $T=0$ limit.  In this limit,
$q\rightarrow 1$, $\lambda$ decouples, and the five other parameters
have a finite value.  The integral in Eq.\ (\ref{E_INT}) can then be
done analytically using a change of variable $m=1-\exp(-\beta y)$ and
keeping only terms with a finite limit when $\beta\rightarrow\infty$,
\begin{equation}
\lim_{\beta\rightarrow\infty} I=\exp(u^2W/4-\Delta u/2)
\mbox{ erfc}\left(\frac{uW-\Delta}{2\sqrt{W}}\right) .
\end{equation}
Shifting $\Delta$
and doing the integral over $u$, we obtain
\begin{eqnarray} \nonumber
\label{E_T0SAD}
\langle\cN_s (f)\rangle&=&\mbox{saddle }
\exp\left\{ N \left[-v\Delta-wW+(f+vW+\Delta/2)^2/W
+G(v+\sqrt{w})\right.\right.\\
&&\left.\left.
+G(v-\sqrt{w})
+\log \mbox{erfc}(-\Delta/2\sqrt{W})\right]\right\} .
\end{eqnarray}
The same result can be obtained directly by counting the average number
of spin configurations $\{\sigma_i\}$ satisfying
\begin{equation}
\forall i\: \sigma_i\sum_{j\neq i}J_{ij}\sigma_j > 0 \mbox{ and }
f=-\sum_{i<j} J_{ij}\sigma_i\sigma_j .
\end{equation}
This computation can be done in rather straightforward way using
Heaviside step functions instead of $\delta$-functions. This approach
is free of the problem of the absolute value of the determinant.

The saddle point equations obtained from Eq.\ (\ref{E_T0SAD}) were
solved numerically to give the shape of the distribution of one-flip
stable configurations as a function of energy density (Fig.\ 2
full-line).  Note that since Eq.\ (\ref{E_T0SAD}) involves the
function $G(x)$ for potentially complex arguments (the complex
conjugate term assures us that the final result will always be real)
one has to choose a sign convention for the square-root terms in the
definition of $G(x)$ (c.f. Eq.\ (\ref{E_GOFX})).  It was chosen such
that the real part of the square-root is always of the same sign as
$v$ in $G(v+\sqrt{w})$; this way the saddle point never has to cross
any cut in the complex plane.

As one follows the saddle point to the rightmost part of
the curve ($e>-0.36$), the argument of the error-function in Eq.\
(\ref{E_T0SAD}) goes through infinity after which all four parameters
become complex. It is not clear if the contour of the integral
computed by the saddle point method can be
deformed to have this complex saddle point has its main
contribution. This issue was not investigated any further for it is of
marginal interest.

In addition, an approximate enumeration of the one-flip stable
configurations for a relatively small ($N=48$) {\sc rom} was done.
Random configurations were generated and then cooled at zero
temperature, the resulting one-flip stable configurations were
compared with previously stored one and were stored if not previously
obtained.  After $4\times10^6$ iterations, the low-energy
configurations were each found about a hundred times while only a few
high-energy ones remained unfound.  The resulting distribution was
binned and its logarithm (divided by 48) is shown as the data points
on Fig.\ 2.  An overall normalization constant (not predicted by the
saddle point method) was added to the simulation data to make them
fall on the predicted curve.  Points on the x-axis correspond to
energy bins where no metastable states were found.  The agreement
between the theory and this simulation is very good.  Similar data
(not shown) for the sine-model also gave very good agreement with the
{\sc rom} prediction.  For {\sc ffm} the Diophantine constraints
(integer spin and integer field) are such that for small $d$ only a
few values of the energy are allowed for a metastable state (e.g.  2
for $d=5$ and 3 for $d=6$). Those constraints might not be so
important for large $d$, but for $d\geq 6$ ($N\geq 64$) an exhaustive
search becomes impossible. Whether the distribution of metastable
states in {\sc ffm} follows that of {\sc rom} remains an open
question.

The point where the curve shown in Fig.\ 3 intersects the x-axis on
the left, corresponds to the minimum energy ($E_{\mbox{\scriptsize
NSol}}$) at which there are on average exponentially many
configurations.  It is almost equal to the approximate ground-state
energy ($E_{\mbox{\scriptsize 1-step}}$) from the one-step replica
breaking solution of this model \cite{MAPARB}.
Careful analysis of the equations leading to these two quantity done
using arbitrary precision arithmetics shows that while they are not
equal they differ by less than 30 parts in a billion.  Precisely,
$E_{\mbox{\scriptsize NSol}}=-0.484119415$ and $E_{\mbox{\scriptsize
1-step}}=-0.484119428$.  The true average ground state energy density
for {\sc rom} is probably well approximated by either of these two
quantities.

\subsection{Finite $T$ Results}

By taking partial derivatives of Eq.\ (\ref{E_NFINAL}), we can write
the number of metastable states in terms of the solution of six
coupled non-linear equations, three of which containing definite
integrals.  A numerical solution of these equations was done by
following the $T=0$ solution of the previous section to finite $T$.
At fixed $T$, varying $u$ traces out the $N^{-1}\log \langle\cN_s
\rangle$ vs $f$ curve.  Typical curves (for $T=0.134$ and $T=0.065$)
are shown on Fig.\ 2 (dash line and dotted line); results for other
temperature are quite similar.  The abrupt stop on the right-hand side
of these curves is due to the divergence of the definite integral in
Eq.\ (\ref{E_NFINAL}), it is possible that the saddle point can be
analytically continued to complex values giving a smooth curve all the
way to zero as in the $T=0$ case.  This issue has not been
investigated since the interesting points lie on the low free energy
end of the curve.

Setting $u=0$ in Eq.\ (\ref{E_NFINAL}) gives the total number of {\sc
tap} solutions. It gives a positive value for $N^{-1}\log \langle\cN_s
\rangle$ starting from 0.2854 at $T=0$ and decreasing all the way to 0
at $T=0.32$.  Above this temperature, the $q=0$ saddle point with
$\log \langle\cN_s \rangle=0$ is the correct one.  The full results
are plotted in Fig.\ 3 (full line).

In the thermodynamics of the model nothing special happens at the
temperature $T=0.32$ at which exponentially many {\sc tap} solutions
appear.  These metastable states have too high free energy to
contribute to the partition function. Instead, if we write
\begin{equation}
e^{-\beta F}=\sum_{\alpha} e^{-\beta F_\alpha}
\approx \int df \langle\cN_s(f)\rangle e^{-N \beta f}
\approx \max_f e^{N A(f)},
\end{equation}
where $\alpha$ labels {\sc tap} solutions and with
\begin{equation}
A(f)\equiv N^{-1}\log \langle\cN_s(f) \rangle-\beta f,
\end{equation}
we find that the solutions which contributes to the free energy are
those for which $A(f)$ is maximum.  They are given by setting
$u=-\beta$ in Eq.\ (\ref{E_NFINAL}). With this substitution, the saddle
point equations admit a simple solution of the form:
\begin{equation}
v=\beta-\frac{\beta q}{2}
\mbox{ , }
w=\frac{\beta^2 q^2}{4}
\mbox{ , }
\Delta=G'(\beta)-G'(\beta(1-q))
\mbox{ , }
W=\frac{\Delta}{\beta}
\mbox{ and }
\lambda=\frac{\beta\Delta}{4}.\label{E_SIMPSAD}
\end{equation}
And the number of contributing {\sc tap} solutions reduces to
\begin{equation}\label{E_MAXQ}
\langle\cN_{cs} \rangle=
\exp\left\{ N\left(\beta f +G(\beta) + \log 2\right)\right\}
\end{equation}
where
\begin{eqnarray} \nonumber
f&=&\beta^{-1}\left[-\log 2-G(\beta(1-q))
-\beta qG'(\beta(1-q))+\beta\Delta\right.\\
&&\qquad\left.
-e^{-\beta\Delta}\int_{-\infty}^{\infty}\frac{dz}{\sqrt{2\pi}}
e^{-\frac{z^2}{2}}
\cosh(\sqrt{2\beta\Delta} z) \log\cosh(\sqrt{2\beta\Delta} z)\right] ,
\end{eqnarray}
with $\Delta(q)$ as above and $q$ determined variationally. The
function $f$ corresponds the free energy of the contributing
solutions. We notice that the logarithm of the number of contributing
solutions, often called {\em complexity}, can be written as
\begin{equation}\label{E_REMI}
\log\langle\cN_{cs} \rangle=\left. \frac{\partial}{\partial m}
\beta F_{\mbox{\scriptsize 1-step}}(\beta,m,q) \right|_{m=1}
\end{equation}
where $F_{\mbox{\scriptsize 1-step}}$ is the free energy computed with
one-step replica symmetry breaking with {\sc rsb} parameter $m$,
$q_0=0$ and extremized over $q_1=q$ (Eq.\ (44) in Ref.
\cite{MAPARB}).  Equality (\ref{E_REMI}) is in fact more general
than what is presented in here, for a more direct derivation and its
physical interpretation see Ref.\ \cite{MONPREP}.

The condition for Eq.\ (\ref{E_MAXQ}) to have a saddle point at a
non-zero value of $q$ is equivalent to the marginality condition
($T<T_{\mbox{\tiny M}}=0.134$). At this temperature, one finds
$q=0.962$ with $N^{-1}\log \langle\cN_{cs}\rangle=0.158$. Monte Carlo
simulations have indicated that all three orthogonal models considered
here ({\sc ffm}, {\sc rom} and the sine model) undergo a dynamical
glassy transition at a temperature equal or very near $T_{\mbox{\tiny
M}}$ \cite{MAPARB,MAPARFF}.

The complexity goes to zero at the replica symmetry breaking
transition $T_{\mbox{\tiny RSB}}=0.065$.  For temperatures below
$T_{\mbox{\tiny RSB}}$, Eq.\ (\ref{E_MAXQ}) gives an unphysical
negative complexity: the above saddle point is no longer valid and the
contributing solutions are those with the smallest free energy (the
leftmost points of the curves of Fig.\ 2). The full curve of
complexity vs.\ temperature is plotted on Fig.\ 3 (dash line).

We finally notice that we naturally have
\begin{equation}
\max_{f} A(f)=\log 2 +G(\beta)=-\beta f_{\mbox{\tiny RS}}.
\end{equation}
In other words even in the region between $T_{\mbox{\tiny RSB}}$ and
$T_{\mbox{\tiny M}}$, where a large number of
solutions contribute to the free energy, the {\sc rs} free energy is
still valid. The decomposition of the {\sc rs} free energy into
contributions from many metastable states below a temperature
``$T_{\mbox{\scriptsize g}}$'' higher than $T_{\mbox{\tiny RSB}}$ has
been discussed in the context of many different models
\cite{KIRTHI,KIRWOL,CRISOM,MONOKA}.

Above $T_{\mbox{\tiny M}}$, one can still find a saddle point of Eq.\
(\ref{E_NFINAL}) with $u=-\beta$, corresponding to the maximum of
$A(f)$, but it is no longer of the simple form (\ref{E_SIMPSAD}).
Numerically one finds that for those temperature, $\max A(f)<-\beta
f_{\mbox{\tiny RS}}$. The metastable states don't contribute to the
free energy.

It may seem surprising that the {\sc rs} solution is still valid
 while a very large number of different states contribute to the free
energy.  The explanation is that their number is so large that if ones
chooses randomly two such states the probability of finding the same
one is zero, therefore the order parameter $P(q)$ only measures the
overlap between different states which is always equal to $q_0=0$.
Only when the complexity ceases to be extensive can the
$P(q)$ become non-trivial, leading to replica symmetry breaking.

\begin{table}
\begin{tabular}{|l|l|}
\hline
\em Temp.&\em Comments \\
\hline
0.40 & Metastable crystalline phase (if present)\\
0.32 & Exponential number of {\sc tap} solutions \\
0.18 & Stability of crystalline solution (if present)\\
0.134 & {\sc tap} solutions start contributing to the free energy,
`marginality condition'\\
&and dynamical phase transition \\
0.065 & Complexity goes to zero and replica symmetry breaking \\
\hline
\end{tabular}
\caption[a]{Interesting temperatures for the orthogonal model, from
text and adapted from Ref.\ \cite{MAPARB} to
present normalizations.}
\end{table}

We now have a simple intuitive scenario for the phase diagram of
models such as {\sc rom} where the order parameter jumps
discontinuously at the transition (see Table 1).  At a relatively high
temperature ($T= 0.32$ in {\sc rom}) the number of metastable states
increases dramatically, but their free energy is so high that they
influence neither the static nor the dynamics.  At $T_{\mbox{\tiny
M}}$, a large number of these states start contributing to the free
energy.  At this temperature, all static thermodynamical quantities
are perfectly regular, but the time scales involved with the dynamics
diverge and the system is no longer able to thermalize.  Phenomena
such as aging start to appear.  At $T_{\mbox{\tiny RSB}}$ the entropy
of the contributing states goes to zero, $P(q)$ becomes non-trivial
and replica symmetry is broken.

\section{Two-Group model}

In this final section we will divert a bit our analysis of the
thermodynamics of the orthogonal model to look at a puzzling
analytical `coincidence.' Indeed, by computing the partition function
for a certain replica-symmetry breaking scheme, we will recover the
expression for the average number of {\sc tap} solution.  The
`two-group model' was first introduced in Ref.\ \cite{BM2G1} as an
attempt to break replica symmetry in {\sc sk}.  It was later noticed
\cite{BM2G2} that the partition function ($\lim_{n\rightarrow
0}\langle Z^{n}_{\mbox{\scriptsize 2G}}\rangle_J$) in this framework
is not equal to unity but instead one has
\begin{equation}
\langle Z^{n}_{\mbox{\scriptsize 2G}}\rangle=
\int df \langle \cN_s(f)\rangle e^{-n\beta Nf},
\end{equation}
and in particular $\lim_{n\rightarrow 0}\langle
Z^{n}_{\mbox{\scriptsize 2G}}\rangle_J$ gives the average number of
{\sc tap} solutions ($\langle\cN_s\rangle_{J}$).  We present here a
general derivation of the partition function for the `two-group
model.' We will assume that the coupling matrix is chosen from a
distribution that obeys a relation such as Eq.\ (\ref{E_GINT}) but we
will not make use of the explicit form of the function $G(x)$. We will
recover the result for {\sc sk} as the special case $G(x)=x^2/4$.

The average partition function for {\sc rom} with $n$ replica is given
by
\begin{equation}
\label{E_PARTF}
\langle Z^{n}\rangle_J =\int d{\bf Q}\:d{\bf\Lambda}\:
\exp\left\{N\left[\tr
G(\beta {\bf Q})-
{\textstyle \frac{1}{2}}\tr({\bf\Lambda Q}) +\log
Z_{\circ}({\bf\Lambda})\right]\right\} ,
\end{equation}
with
\begin{equation}
Z_{\circ}({\bf\Lambda})\equiv \sum_{\sigma^a}
\exp\left\{\sum_{a<b}\Lambda_{ab}\sigma^a\sigma^b\right\} .
\end{equation}

The two-group ansatz correspond to writing the matrix $\bf\Lambda$ and
$\bf Q$ as made up of two diagonal blocks of size $m\times m$ and
$(n-m)\times (n-m)$ and equal elements outside those blocks. We
specialize right away to the case $n=0$ which will make some of our
formulae look a bit strange but will make the notation more compact.
The following derivation can be extended without much difficulties to
the $n\neq 0$ case. We define
\begin{equation}
{\bf \Lambda}=
\left(\mbox{\rule[-1.5ex]{0cm}{3ex}}\right.
    \overbrace{
      \begin{array}{c}
         u_1\\
         u
       \end{array}
      }^{m}
     \:
    \overbrace{
      \begin{array}{c}
         u\\
         u_2
       \end{array}
      }^{-m}
      \left.\mbox{\rule[-1.5ex]{0cm}{3ex}}\right)
\mbox{ and }
{\bf Q}=
\left(\mbox{\rule[-1.5ex]{0cm}{3ex}}\right.
    \overbrace{
      \begin{array}{c}
         q_1\\
         q
       \end{array}
      }^{m}
     \:
    \overbrace{
      \begin{array}{c}
         q\\
         q_2
       \end{array}
      }^{-m}
      \left.\mbox{\rule[-1.5ex]{0cm}{3ex}}\right) ,
\end{equation}
with $\Lambda_{aa}=0$ and $Q_{aa}=1$ and with the elements
parameterized as
\begin{equation}
u_1=u+t/m+r/m^2 \; , \; u_2=u-t/m+r/m^2
\end{equation}
\begin{equation}
q_1=q+\alpha/m+\gamma/m^2 \; , \; q_2=q-\alpha/m+\gamma/m^2 ,
\end{equation}
where $m$ will be taken to infinity.  Having defined the two-group
ansatz, we can now compute the different parts of Eq.\
(\ref{E_PARTF});
\begin{equation}
\label{E_TG1}
\tr({\bf\Lambda Q})=2(\alpha t+qr+\gamma u) - 2(\alpha u+q t).
\end{equation}
To compute $\tr G(\beta {\bf Q})$ we find the different eigenvalues of
the matrix $\bf Q$ and their multiplicity, leading us to
\begin{eqnarray} \nonumber
\tr G(\beta {\bf Q})&=& \lim_{m\rightarrow \infty}\:
(m-1)G(\beta(1-q_1))-(m+1)G(\beta(1-q_2))+G(\beta q^+)+G(\beta q^-)\\
&=&-2G(\beta(1-q))-2\alpha\beta G'(\beta(1-q))+G(\beta q^+)
+G(\beta q^-) ,
\label{E_TG2}
\end{eqnarray}
where
\begin{equation}
q^{\pm}=q-\alpha-1\pm\sqrt{2q(\gamma-\alpha)} .
\end{equation}
The $Z_{\circ}({\bf\Lambda})$ term requires more work,
\begin{equation}
Z_{\circ}({\bf\Lambda})=\lim_{m\rightarrow \infty}\: e^{-t}
\sum_{\sigma^a}\exp\left\{\frac{1}{2}
\left[u S^2
+\left(\frac{t}{m}+\frac{r}{m^2}\right)S_+^2
+\left(-\frac{t}{m}+\frac{r}{m^2}\right)S_-^2
\right]\right\} ,
\end{equation}
where
\begin{equation}
S=\sum \sigma_a\:,\:
S_+=\sum_{a\leq m} \sigma_a\:\mbox{and}\:
S_-=\sum_{a> m} \sigma_a .
\end{equation}
If we make three Hubbard-Stratonovich transformations to linearize the
quadratic terms and do the sum over spins, we find
\begin{eqnarray} \nonumber
Z_{\circ}({\bf\Lambda})&=&\lim_{m\rightarrow \infty}\: e^{-t}
\int_{-\infty}^{\infty} \frac{dh}{\sqrt{2\pi u}}
\int_{-\infty}^{\infty}\int_{-\infty}^{\infty}
\frac{m dh_+\, dh_-}{2\pi t}
\exp\left\{\frac{1}{2}\left[-\frac{h^2}{u}+
\frac{r(h_+^2+h_-^2)}{t^2}\right]\right.\\
&&+m\left.\left[\log\cosh(h+h_+)-\log\cosh(h+h_-)
-\frac{h_+^2}{2 t}
+\frac{h_-^2}{2 t}\right]\right\} ,
\end{eqnarray}
where we have put in evidence the terms with a factor of $m$. We can
now use the saddle point method to compute the $m\rightarrow\infty$
limit of the last expression. To leading order in $m$, the saddle
point equations for $h_+$ and $h_-$ are the same.  They do differ by a
term of order $1/m$ but this perturbation only contributes to the
saddle point at order $1/m$ (it does not contribute at order one
because the derivative of the order $m$ term vanishes).  At the saddle
point, one has $h_+=h_-=h_{\circ}$ which solves
\begin{equation}
E(h,h_{\circ})\equiv\tanh(h+h_{\circ})-\frac{h_{\circ}}{t}=0 .
\end{equation}
The order $m$ contributions to the saddle point of $h_+$ and $h_-$
exactly cancel each other leaving us with a finite result as
$m\rightarrow\infty$. Including Gaussian fluctuations, we find
\begin{equation}
Z_{\circ}({\bf\Lambda})=e^{-t}
\int_{-\infty}^{\infty} \frac{dh}{t\sqrt{2\pi u}}
\left(\frac{\partial E}{\partial h_{\circ}}\right)^{-1}
\exp\left[\frac{1}{2}\left(-\frac{h^2}{u}
+\frac{2 r h^2_{\circ}}{t^2}\right)\right] .
\end{equation}
Using the implicit function theorem,
\begin{equation}
dh\left(\frac{\partial E}{\partial h_{\circ}}\right)^{-1}
=dh_{\circ}\left(\frac{\partial E}{\partial h}\right)^{-1}
=dh_{\circ}\left(1-\frac{h_{\circ}^2}{t^2}\right)^{-1} ,
\end{equation}
we can change the integration variable from $h$ to $h_{\circ}$ which
we rescale\footnote{Note that we call $m$ the rescaled integration
variable $h_{\circ}/t$ to make the connection with Eq.\
(\ref{E_NFINAL}), it is not to be confused with the replica-symmetry
breaking parameter $m$ which as been taken to infinity.}  to find
\begin{equation}
Z_{\circ}({\bf\Lambda})=e^{-t}
\int_{-1}^{1} \frac{dm}{\sqrt{2\pi u}}
\left(\frac{1}{1-m^2}\right)
\exp\left[-\frac{\left(\tanh^{-1}m-t m\right)^2}{2 u}
+r m^2\right] .
\label{E_TG3}
\end{equation}

Finally if we collect the results from Eqs.\
(\ref{E_TG1}, \ref{E_TG2}, \ref{E_TG3}), make the following change of
variables
\begin{equation}
\Delta=t/\beta \:,\: W=u/2\beta^2
\:,\:\lambda=r\:,\:
w=\beta^2[2q(\gamma-\alpha)]
\mbox{ and }
v=\beta(\alpha+1-q) ,
\end{equation}
and extremize over the six parameters, we recover $\langle\cN_s\rangle$
given by Eq.\ (\ref{E_NFINAL}) with $u=0$ where the saddle point with
$B=0$ and $b=1-q$ has been chosen. Had we not set $n=0$, we would have
recovered the r.h.s.\ of Eq.\ (\ref{E_NFINAL}) with $u=-n\beta$ and
without the $-uf$ term. In other words $\langle
Z^{n}_{\mbox{\scriptsize 2G}}\rangle$ is the Laplace transform of
$\langle\cN_s(f)\rangle$ with $-n\beta N$ conjugate to $f$.

\subsection*{Acknowledgments}
We thank J. Kurchan, E. Marinari, R. Monasson, F. Ritort and M. Wexler
for useful discussions.

\appendix
\section*{Appendix}
We will now proceed to show that the determinant of the matrix $A$ and
the term $\sim J_{ij}x_i m_j$ (c.f. Eq.\ (\ref{E_NSOL1})) can be
averaged separately. Had we kept them together, introduced Grassman
variables and integrate over the couplings, we would have obtained a
term like this
\begin{equation}
\exp\left\{N\tr G\left[
\beta\frac{{\bf x}\otimes{\bf m}+{\bf m}\otimes{\bf x}
+\bthb\otimes\bth-\bth\otimes\bthb}{N}
\right]\right\}
\end{equation}
Therefore we need to show that the cross-terms in the Taylor expansion
of $\tr G[\ldots]$ are irrelevant in the large $N$ limit. A typical
cross-term is
\begin{equation}
\frac{1}{N^{n-1}}\sum_{\{i_k\}}  x_{i_1}m_{i_2}x_{i_2}m_{i_3}
\theta_{i_3}
\thb_{i_4}\theta_{i_4}\thb_{i_5}m_{i_5}x_{i_6}\ldots\thb_{i_n}
\theta_{i_1}
\end{equation}
We can consider those terms as a perturbation about the fermionic
Gaussian integral (\ref{E_FERM}).  We need to compute all connected
diagrams generated by these new $\theta$ and $\thb$ vertices,
contracting them with the Gaussian propagator $(a_i+\beta
R)^{-1}\delta_{ij}$.  The point is that since the propagator is
diagonal, all such contractions will be at most of order 1 while the
logarithm of the unperturbed result (Eq.\ (\ref{E_DETFINAL})) is of
order $N$.  The same is true of the term we dropped in writing Eq.\
(\ref{E_ADEF}).


\begin{thebibliography}{99}

\bibitem{BOUMEZ}
J.-P. Bouchaud and M. M\'ezard, Self Induced Quenched Disorder:
A Model for the Glass Transition,
{\em J. Phys. I France }{\bf 4} (1994) 1109.

\bibitem{MAPARA}
 E. Marinari, G. Parisi and F. Ritort,
 Replica Field Theory for Deterministic Models: Binary Sequences
 with Low Autocorrelation,
{\em J. Phys. A (Math. Gen.)} {\bf 27} (1994) 7615.

\bibitem{AMORPH}
G. Parisi, Some Applications of Field Theory to Amorphous Systems, in
    {\em Quantum  Field  Theory  and  Quantum  Statistics.
    Essays in honour of the
    sixtieth birthday of E. S. Fradkin},  edited by
    I. A. Batalin, C. J.  Isham
    and G. A. Vilkovisky, Adam Hilger (Bristol) 1987.

\bibitem{DISSIM}
G.Parisi, {\em Field Theory, Disorder and Simulations,}
World Scientific
(Singapore) 1992.

\bibitem{DOTPAR}
G. Parisi and V. Dotskenko, Random Magnetic Fields and Instantons in
Replica Space,
{\em J. Phys. A (Math. Gen.)} {\bf 25} (1992) 3143.

\bibitem{MAPARB}
  E. Marinari, G. Parisi and F. Ritort,
Replica Field Theory for Deterministic Models (II): A Non-Random
  Spin Glass with Glassy Behavior,
{\em J. Phys.  A (Math. Gen.) }{\bf 27} (1994) 7647.

\bibitem{TOUL}
G. Toulouse,
Theory of Frustration Effect in Spin Glasses: I,
{\em Comm. Phys. }{\bf 2} (1977) 115.

\bibitem{DPTV1}
  B. Derrida, Y. Pomeau, G. Toulouse and J. Vannimenus,
Fully Frustrated Simple Cubic Lattices and the Overblocking
       Effect,
{\em   J. Physique }{\bf 40} (1979) 617.

\bibitem{DPTV2}
  B. Derrida, Y. Pomeau, G. Toulouse and J. Vannimenus,
Fully Frustrated Simple Cubic Lattices and Phase Transitions,
{\em   J. Physique }{\bf 41} (1980) 213.

\bibitem{MAPARFF}
  E. Marinari, G. Parisi and F. Ritort,
The Fully Frustrated Hypercubic Model is Glassy and
       Aging at Large $D$,
{\em J. Phys.  A (Math. Gen.) }{\bf 28} (1995) 327.

\bibitem{PLEFKA}
T. Plefka,
Convergence Condition of the {\sc tap} Equation for the Infinite-Range
Ising Spin Glass Model,
{\em J. Phys.  A (Math. Gen.)} {\bf 15} (1982) 1971.

\bibitem{YEDGE2}
  A. Georges and J. S. Yedidia,
How to Expand around Mean-Field Theory Using High-Temperature
  Expansions,
{\em   J. Phys.  A (Math. Gen.) }{\bf 24} (1991) 2173.

\bibitem{YEDGE1}
  J. S. Yedidia and A. Georges,
The Fully Frustrated Ising Model in Infinite Dimensions,
{\em   J. Phys.  A (Math. Gen.) }{\bf 23} (1990) 2165.

\bibitem{KUPAVI}
J. Kurchan, G. Parisi and M. A. Virasoro,
Barriers and Metastable States as Saddle Points in the Replica
Approach,
{\em J. Phys. I France }{\bf 3} (1993) 1819

\bibitem{TAP}
D. J. Thouless, P. W. Anderson and R. G. Palmer,
Solution of `Solvable Model of a Spin Glass',
{\em Phil. Mag. }{\bf 35} (1977) 593.

\bibitem{BM}
A. J. Bray and M. A. Moore,
Metastable States in Spin Glasses,
{\em J. Phys. C (Solid St. Phys.)} {\bf 13} (1980) L469.

\bibitem{RIEGER}
H. Rieger,
The Number of Solutions of the Thouless-Anderson-Palmer Equations for
the $p$-Spin-Interaction Spin Glasses,
{\em Phys. Rev. }{\bf B 46} (1992) 14655.

\bibitem{PARSOU}
G. Parisi and N. Sourlas,
Supersymmetric Field Theories and Stochastic Differential Equations,
{\em Nucl. Phys.} {\bf B 206} (1982) 321.

\bibitem{KURABS}
J. Kurchan,
Replica Trick to Calculate Means of Absolute Values: Applications to
Stochastic Equations,
{\em J. Phys. A }(Math. Gen.) {\bf 24} (1991) 4969.

\bibitem{MONPREP}
R. Monasson, in preparation.

\bibitem{KIRTHI}
T. R. Kirkpatrick and D. Thirumalai, $p$-spin Interaction
Spin-Glass Models: Connections with the Structural Glass Problem,
{\em Phys. Rev. B. }{\bf 36} (1987) 5388.

\bibitem{KIRWOL}
T. R. Kirkpatrick and P. G. Wolynes,
Stable and Metastable States in Mean-Field Potts and Structural Glasses,
{\em Phys. Rev. B. }{\bf 36} (1987) 8552.

\bibitem{CRISOM}
A. Crisanti and H.-J. Sommers,
{\em On The TAP Approach to the Spherical $p$-spin SG Model}, preprint
TNT-94-5, cond-mat/9406051, submitted to {\em J. Phys. I France}..

\bibitem{MONOKA}
R. Monasson and  D. O'Kane,
Domains of Solutions and Replica Symmetry Breaking in Multilayer
Neural Networks,
{\em  Europhys. Lett. }{\bf 27}  (1994) 85.

\bibitem{BM2G1}
A. J. Bray and M. A. Moore,
Replica-Symmetry Breaking in Spin-Glass Theories,
{\em Phys. Rev. Lett. }{\bf 41} (1978) 1068.

\bibitem{BM2G2}
A. J. Bray and M. A. Moore,
Broken Replica Symmetry and Metastable States in  Spin Glasses,
{\em J. Phys. C (Solid St. Phys.) }{\bf 13} (1980) L907.

\end{thebibliography}
\end{document}